\documentclass[12pt]{article}

\usepackage{graphicx,tabularx}
\usepackage{bm}
\usepackage{epsfig}
\usepackage{citesort}

\newcommand{\beq}{\begin{equation}}
\newcommand{\eeq}{\end{equation}}
\newcommand{\beqn}{\begin{eqnarray}}
\newcommand{\eeqn}{\end{eqnarray}}
\newcommand{\iden}{\leavevmode\hbox{\small1\normalsize\kern-.33em1}}
\renewcommand{\theequation}{\arabic{equation}}

\def\ds{\displaystyle}
\def \bcen     {\begin{center}}
\def \ecen     {\end{center}}
\def \beqa     {\begin{eqnarray}}
\def \eeqa     {\end{eqnarray}}

\def \gmtwo     {$(g - 2)_\mu$ }
\def \mutoeg    {$\mu \to e \gamma$ }
\def \tautomug  {$\tau \to \mu \gamma$ }
\def \foftga    {$f_1 \to f_2 \gamma$ }
\def \amu       {a_\mu}
\def \vpmns     {V_{PMNS}}

\begin{document}

\renewcommand{\theequation}{\thesection.\arabic{equation}}
\renewcommand{\thefootnote}{\fnsymbol{footnote}}

\begin{titlepage}


\begin{flushright}
KEK-TH-1129, \\
LYCEN 2006-23, \\
hep-ph/0612327
\end{flushright}
\vskip 1cm

\bcen

{\bf \Large Lepton flavour violation in The Little Higgs model}
\vskip 2.5cm
{\sffamily \bf 
S. Rai Choudhury$^1$ \footnote{src@physics.du.ac.in},
A. S. Cornell$^2$ \footnote{cornell@ipnl.in2p3.fr}, 
A. Deandrea$^2$ \footnote{deandrea@ipnl.in2p3.fr}, 
Naveen Gaur$^3$ \footnote{naveen@post.kek.jp}, 
A. Goyal$^4$ \footnote{agoyal@iucaa.ernet.in} }

\vskip 1.5cm
{\sl 
$^1$ Center for Theoretical Physics, Jamia Millia University, Delhi - 110025,
India,  \\ \vspace{.4cm}
$^2$ Universit\'e de Lyon 1, Institut de Physique Nucl\'eaire, \\
4 rue E. Fermi, F-69622 Villeurbanne Cedex, France, \\ \vspace{.4cm}
$^3$ Theory Division, KEK, 1-1 Oho, Tsukuba, Ibaraki 305-0801, Japan
\\ \vspace{.4cm}
$^4$ Dept. of Physics \& Astrophysics, University of Delhi, Delhi - 110 007, India }

\ecen
\vskip 1cm

\begin{abstract}
\noindent Little Higgs models with T-parity have a new source of lepton flavour
violation. In this paper we consider the  anomalous magnetic moment of
the muon \gmtwo and the lepton flavour violating decays \mutoeg and
\tautomug in Little Higgs model with T-parity \cite{Goyal:2006vq}.
Our results shows that present experimental constraints of \mutoeg is
much more useful to constrain the new sources of flavour violation
which are present in T-parity models. 
\end{abstract}
\end{titlepage}




\section{Introduction}

\par Electroweak precision data suggests that the physics of
electroweak symmetry breaking is weakly coupled, therefore, in order
to have a natural theory, the Higgs mass needs to be protected from
radiative corrections. As a means to solve this problem a number of
extensions of the Standard Model (SM) have been proposed. In the
effective theory approach the collective symmetry breaking mechanism
of the Little Higgs (LH) models is an interesting possibility
\cite{lh}. For earlier attempts to solve these issues see reference
\cite{earlyHiggs}. Note that a detailed review of LH models can be
found in reference \cite{Schmaltz:2005ky} (see also chapter 7 of
reference \cite{Accomando:2006ga}). Also, as the electroweak sector of
the SM has been tested to a very high accuracy, an important test of
the validity LH models through a comparison with precision data (for
reviews treating this subject see references
\cite{Perelstein:2005ka,Chen:2006dy}). Note that there exists many
studies in the literature concerning the LH model and its implications
to electroweak corrections (see for example reference \cite{LHEW}) and
flavour physics both in hadronic \cite{Buras:2004kq} and leptonic
sector \cite{Choudhury:2005jh}.  
 
\par The most serious constraints result from the tree-level
corrections to precision electroweak observable due to the exchanges
of the additional heavy gauge bosons present in the model, as well as
from the small but non-vanishing vev of an additional weak-triplet
scalar field. As a result, the fine-tuning of the Higgs boson mass is
re-introduced. In order to forbid dangerous tree-level contributions
to the electroweak observable, and avoid the appearance of a new fine
tuning between the electroweak scale and the scale of the model,
several new variants were proposed 
\cite{Chang:2003un}. Particularly interesting is the implementation of
a $Z_2$ symmetry called T-parity \cite{Low:2004xc}. T-parity
explicitly forbids tree-level contribution from the new heavy gauge
bosons to the observable involving only SM particles as external
states. It also forbids the interactions that induce triplet vev
contributions. In T-parity symmetric LH models (LHT), corrections to
precision electroweak observables are generated exclusively at loop
level. Due to T-parity the lightest T-odd particle becomes stable.
Since this lightest T-odd particle is electrically and colour neutral
with ${\cal O}(100)$GeV it could be a candidate for dark matter
\cite{Cheng:2005as,Asano:2006nr}.  

\par LHT predict heavy T-odd gauge bosons which are the T-partners of
the SM gauge boson and also heavy T-odd $SU(2)$ doublet fermions. This
structure is unique to LHT, and as the new particle masses can be
relatively low, the next generation of colliders such as the Large
Hadron Collider (LHC) has the great potential to directly produce the
T-partners of the SM particles \cite{phenoLHT}. The peculiar structure
of these models can also be tested using  precision data, present and
future colliders \cite{Hubisz:2004ft}. In particular the flavour
structure of the model can be constrained both  
in the quark sector \cite{Blanke:2006eb} and in the lepton sector. 

The neutrino oscillation data from experiments is proving the
existence of small neutrino mass and large neutrino flavour mixing. If
the small neutrino mass as hinted by experiments is the only source of
lepton flavour violation (LFV), then the LFV processes like \mutoeg,
$\tau \to \mu \mu \mu$ etc. would be heavily suppressed because of
lepton sector GIM. The presence of new sources of LFV can enhance
these processes to the level of present experimental limit. Little
Higgs model with T-parity have a possible source of lepton flavour
violation. In
the following we shall concentrate on the mirror lepton sector, and on
the interplay with the heavy T-odd gauge bosons sector of the Little
Higgs model with T-parity,
by studying the anomalous magnetic moment of the muon \gmtwo and the
lepton family violating decays \mutoeg and \tautomug, $\tau \to e
\gamma$, $\tau \to \mu \mu \mu$, $\tau \to \mu 
M$ (semi-leptonic decay) etc..


\section{Littlest Higgs model with T-parity}

\par In this section we shall briefly review the Littlest Higgs model
with T-parity of reference \cite{Low:2004xc}, in order to present our
notation. We follow here, for the leptons, a notation similar to the
one used by Buras {\it et al.}\cite{Blanke:2006eb} in the analysis of
non-minimal flavour violating interactions in the quark sector for
LHT. Note that the model is a non-linear chiral-type Lagrangian based
on the coset $SU(5)/SO(5)$. 

\par The first stage of symmetry breaking is at a scale $f$ in the TeV
range, and is due to the vacuum expectation value (vev) of an $SU(5)$
symmetric matrix $\Sigma$, that is: 
\beq
\Sigma_0 = \left( \begin{array}{ccc}
0 & 0 & \iden \\
0 & 1 & 0 \\
\iden & 0 & 0 
\end{array} \right) , 
\label{eq:sec2:1}
\eeq
where $\iden$ is the $2\times 2$ identity matrix. This breaking
simultaneously breaks the gauge group to an $SU(2) \times U(1)$
subgroup, which is identified with the SM group. The origin of this
symmetry breaking is not specified in the model but merely
imposed. Therefore LHT are effective theories, valid up to a scale
$\Lambda\sim 4\pi f$, as can be established in analogy with similar
arguments in chiral Lagrangians. The generators, $T^a$, of the
unbroken $SO(5)$ symmetry, are those which satisfy the relation
$T^a\Sigma_0+\Sigma_0 (T^a)^T=0$. The broken generators, $X^a$, of
$SU(5)/SO(5)$, satisfy the relation $X^a\Sigma_0-\Sigma_0
(X^a)^T=0$. The $SU(5)/SO(5)$ breaking gives rise to 14
Nambu-Goldstone bosons; four of the fourteen Goldstone bosons are
absorbed by the broken gauge generators, and the remaining ten
Goldstones are parameterized as:   
\beq
\Pi = \left(
\begin{array}{ccc}
            & h^\dagger/\sqrt{2}  &  \Phi^\dagger   \\
h/\sqrt{2}  &   & h^*/\sqrt{2}                \\
\Phi        & h^T/\sqrt{2}  &    
\end{array}
\right) ,
\eeq
where $h$ is the SM Higgs doublet and $\Phi$ is a complex $SU(2)$
triplet:   
\beq
\Phi = \left(
\begin{array}{cc}
\Phi^{++} & \Phi^+/\sqrt{2}     \\
\Phi^+/\sqrt{2}  & \Phi^0
\end{array}
\right) .
\eeq
The second stage of symmetry breaking takes place as in the SM via the
usual Higgs mechanism, at a scale $v=256$ GeV. 

\par The effective theory at low energy is described by a chiral-type
Lagrangian with the appropriate gauging (a $[SU(2)\times U(1)]^2$
subgroup of the global $SU(5)$ symmetry is gauged). T-parity exchanges
the two $SU(2)\times U(1)$ factors. The symmetric tensor describing
the low energy theory is: 
\begin{equation}
\Sigma=e^{i\Pi/f}\Sigma_0 e^{i\Pi^T/f} = e^{2i\Pi/f}\Sigma_0 =
\Sigma_0 + \frac{2 i}{f} \Pi \Sigma_0 + {\cal O}(1/f^2) \; , 
\end{equation}
where $f$ is the scale of symmetry breaking we have just described;
similar to $f_\pi$ in the case of chiral Lagrangians. The kinetic term
for the $\Sigma$ field can be written as: 
\beq 
{\cal{L}}_{kin} = \ds \frac{f^2}{8} \mathrm{Tr} \left\{D_{\mu} \Sigma (
D^{\mu} \Sigma)^{\dagger} \right\} , 
\eeq
where
\beq
D_{\mu} \Sigma = \partial_{\mu} \Sigma - i \Sigma_j \left[ g_j W_j^a
(Q_j^a \Sigma + \Sigma Q_j^{aT}) + g'_j B_j (Y_j \Sigma + \Sigma Y_j )
\right] .  
\eeq
In the above $j = 1, 2$, the $Q_j$ and $Y_j$ are the gauged
generators, $B_j$ and $W_j^a$ are the $U(1)_j$ and $SU(2)_j$ gauge
fields, respectively, and $g_j$ and $g'_j$ are the corresponding
coupling constants.    

\subsection{Gauge bosons sector}

\par In the gauge boson sector the gauge boson eigenstates are
identified as: 
\begin{eqnarray}
W^a_L=\frac{W^a_1+W^a_2}{\sqrt{2}}\,,\qquad
B_L=\frac{B_1+B_2}{\sqrt{2}},\\
W^a_H=\frac{W^a_1-W^a_2}{\sqrt{2}}\,,\qquad
B_H=\frac{B_1-B_2}{\sqrt{2}}, 
\end{eqnarray}
where $L$ refers to the light (and T-even) states and $H$ the heavy
(and T-odd) states. The mass eigenstates are then given, at
${\cal{O}}(v^2/f^2)$, by the following combinations of gauge boson
eigenstates: 
\begin{eqnarray}
W^\pm_L&=&\frac{W^1_L \mp i W^2_L}{\sqrt{2}}\,,\qquad
Z_L=\cos\theta_W W^3_L-\sin\theta_WB_L\,,\qquad
A_L=\sin\theta_W W^3_L+\cos\theta_WB_L\,,\\
W^\pm_H&=&\frac{W^1_H \mp i W^2_H}{\sqrt{2}}\,,\qquad
Z_H=W^3_H+x_H\frac{v^2}{f^2} B_H\,,\qquad
A_H=-x_H\frac{v^2}{f^2} W^3_H+B_H\,,
\end{eqnarray}
where $\theta_W$ is the weak mixing angle and $x_H =
5gg'/4(5g^2-g'^2)$ with $g$ and $g'$ being, respectively, the $SU(2)$
and $U(1)$ gauge couplings. The gauge boson masses are then given at
${\cal{O}}(v^2/f^2)$ by: 
\begin{equation}
M_{W_H}= f g\left(1-\frac{v^2}{8f^2}\right),\qquad
M_{Z_H}\equiv M_{W_H} \,,\qquad 
M_{A_H}=\frac{f g'}{\sqrt{5}}\left(1-\frac{5v^2}{8f^2}\right) . 
\end{equation}
The masses of the T-even gauge bosons are zero after the first stage
of symmetry breaking and obtain a mass only through the second
breaking, their masses being: 
\begin{equation}
M_{W_L}=\frac{gv}{2}\left(1-\frac{v^2}{12f^2}\right),\quad
M_{Z_L}=\frac{gv}{2\cos\theta_W}\left(1-\frac{v^2}{12f^2}\right),\quad
M_{A_L}=0\,. 
\end{equation}

\subsection{Mirror fermion sector and mixing}

\par For each SM $SU(2)_L$ doublet, a doublet under $SU(2)_1$ and
another under $SU(2)_2$ are introduced. The T-parity even linear
combination is associated with the SM $SU(2)_L$ doublet, while the
T-odd combination is given a mass of order of the scale $f$. This is
required for a consistent implementation of T-parity in the fermion
sector \cite{Low:2004xc}. The fermion doublets are embedded into the
following incomplete representations of $SU(5)$, and introduce a
right-handed $SO(5)$ multiplet $\Psi_R$:  
\beq
\Psi_1 = \left(\begin{array}{c} i \psi_1 \\ 0
    \\0 \end{array}\right),\qquad 
\Psi_2 = \left(\begin{array}{c} 0 \\ 0 \\
    i\psi_2 \end{array}\right),\qquad 
\Psi_R = \left(\begin{array}{c} \tilde{\psi}_R \\ \chi_R\\
    \psi_R \end{array}\right), 
\eeq
with
\beq
\psi_i=-\sigma^2 f_i=
-\sigma^2\begin{pmatrix}u_i\\d_i\end{pmatrix}\,,\qquad
\psi_R=-i\sigma^2\begin{pmatrix}u_{HR}\\d_{HR}\end{pmatrix}
\eeq
and $(i=1,2)$. Under T-parity these fields transform in the following
way: 
\beq
\Psi_1\mapsto
-\Sigma_0\Psi_2\,,\qquad \Psi_2\mapsto -\Sigma_0\Psi_1\,,\qquad
\Psi_R\mapsto -\Psi_R\, ,
\eeq
and the T-parity eigenstates of the fermion doublets are:
\beq
f_L=\frac{f_1-f_2}{\sqrt{2}}\,,\qquad
f_H=\frac{f_1+f_2}{\sqrt{2}}\,.
\eeq
$f_L$ are the left-handed SM fermion doublets (T-even), and $f_H$ are
the left-handed mirror fermion doublets (T-odd). The right-handed
mirror fermion doublet is given by $\psi_R$. The mirror fermions
obtain a mass of the order of the scale $f$ \cite{phenoLHT}: 
\beqn 
m^d_{Hi}&=&\sqrt{2}\kappa_i f \equiv m_{Hi} , \\
m^u_{Hi}&=&\sqrt{2}\kappa_i f\left(1-\frac{v^2}{8f^2}\right)=
m_{Hi} \left(1-\frac{v^2}{8f^2}\right)\,,
\eeqn
where $\kappa_i$ are the eigenvalues of the mass matrix $\kappa$. The
additional fermions $\tilde\psi_R$ and $\chi_R$ can be given large
Dirac masses, and we assume that they are decoupled from the theory. 

\par In a similar way to what happens for standard fermions, the
mirror sector has weak mixing, parameterised by unitary mixing
matrices; two for mirror leptons and two for mirror quarks: 
\beq
V_{H\ell}\,,\quad V_{H\nu}\,,\quad V_{Hu}\,,\quad V_{Hd}\,, 
\eeq
which satisfy the following physical constraints:
\beq
V_{H\nu}^\dagger V_{H\ell}=V_{PMNS}\,,\qquad  V_{Hu}^\dagger
V_{Hd}=V_{CKM}\,. 
\eeq
Furthermore, this implies that one can not turn off the new mixing
effects, except with a universal degenerate mass spectrum for the
T-odd doublets. In the following we shall set the Majorana phases of
$V_{PMNS}$ to zero, as no Majorana mass term has been introduced for
the right-handed neutrinos. The mixing in the lepton sector will be
the main focus of our phenomenological analysis. Note that a detailed
discussion of the parameterisation of the mirror lepton mixing
matrices is given in section IV.


\section{Analytical results of \gmtwo and  \mutoeg} 

\par In this section we shall summarize our analytic results for the
anomalous magnetic moment of the muon \gmtwo and the two lepton family
violating decays \mutoeg and \tautomug from the Littlest Higgs model
with T-parity.  

\subsection{Anomalous magnetic moment of the muon}

\par
\begin{figure}[htb]
\vskip .5cm
\bcen
\epsfig{file=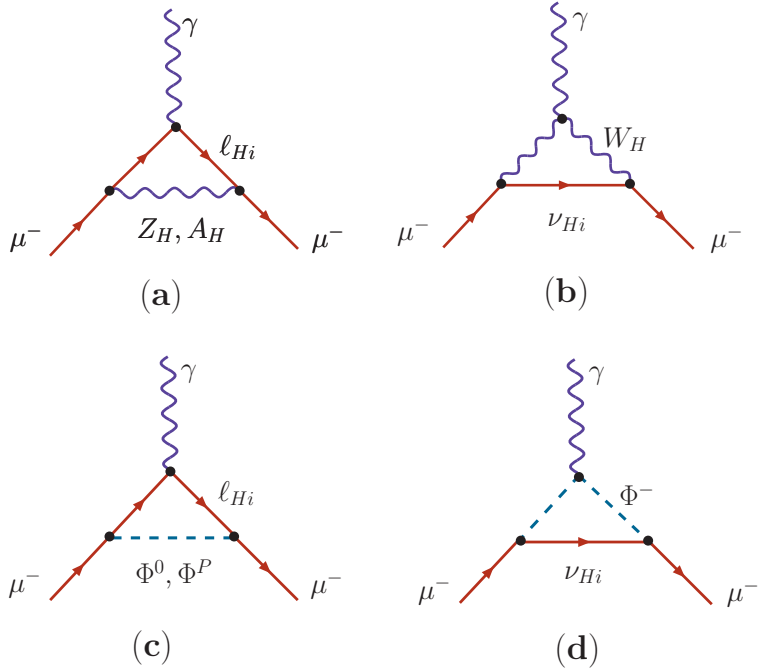,width=.6\textwidth}
\ecen
\caption{Diagrams of the additional contributions to the \gmtwo arising from LHT}
\label{fig:1}
\end{figure}

 For the \gmtwo in LHT we have the following additional contributions:
\begin{itemize}
\item{} The $W^\pm_H$ contribution. 
\item{} The $Z_H$ contribution
\item{} The $A_H$ contribution.
\item{} And the $\Phi$ (triplet Higgs) contribution. 
\end{itemize}
As such, the additional diagrams which will contribute to \gmtwo (at
one loop level) are given in Fig (\ref{fig:1}), and where the
contribution to the $\amu$ due to the new particles can be written as:
\beq
\amu^{LH} = \amu(W_H) + \amu(A_H) + \amu(Z_H) + \amu(\Phi) . 
\label{eq:4}
\eeq

\par Note that the contributions due to triplet Higgs ($\Phi$) can be
neglected, as their coupling to SM fermions and T-odd fermions is of
order $(v/f)^2$ (see Appendix A of reference \cite{Blanke:2006eb}). As
such, the diagrams Fig (\ref{fig:1} c), and (\ref{fig:1} d) will give
contributions at the $(v/f)^4$ level (as these diagrams have two
vertices involving the triplet Higgs). In order $(v/f)^2$ calculations
we can therefore neglect any such contributions. This means we shall
only consider contributions due to heavy gauge bosons given by
diagrams Fig (\ref{fig:1} a) and (\ref{fig:1} b). 

\par The various contributions due to gauge bosons in the unitary
gauge are: 
\beqa
\amu(W_H) &=& - \frac{g^2}{32 \pi^2} \frac{m_\mu^2}{M_{W_H}^2}
\sum_{i=1,3} \left( V_{H\nu} \right)_{2i}^* \left( V_{H\nu}
\right)_{2i} 
F_{W_H} (x_i) \quad , \quad \mathrm{where} \quad 
x_i = \left( \frac{M_{\nu_{H_i}}}{M_{W_H}}\right)^2 , 
\label{eq:5}  \\
\amu(Z_H) &=& \frac{g^2}{32 \pi^2} \frac{m_\mu^2}{M_{Z_H}^2}
\sum_{i=1,3} \left( V_{H\ell} \right)_{2i}^* \left( V_{H\ell}
\right)_{2i} F_{Z_H} (y_i) 
\quad , \quad \mathrm{where} \quad 
y_i = \left( \frac{M_{\ell_{H_i}}}{M_{Z_H}}\right)^2 ,   
\label{eq:6} \\ 
\amu(A_H) &=& 
\frac{g'^2}{800 \pi^2} \frac{m_\mu^2}{M_{A_H}^2}
\sum_{i=1,3} \left( V_{H\ell} \right)_{2i}^* \left( V_{H\ell}
\right)_{2i} F_{Z_H} (z_i) 
\quad , \quad \mathrm{where} \quad 
z_i = \left( \frac{M_{\ell_{H_i}}}{M_{A_H}}\right)^2 , 
\label{eq:7}
\eeqa
and where the functions $F_X (x)$ have been defined in Appendix \ref{GminusTwo}.

\par In the next section we shall generate plots of $a_{\mu}^{LH}$ for
various values of $f$ and mirror lepton masses. 


\subsection{Lepton family violating decays}

We shall now consider the lepton family violating decays of the form
$f_1(p_1) \to f_2(p_2) \gamma(q)$, with $q = p_1 - p_2$. In these
calculations we shall take the fermions $f_1$ and $f_2$ as having
masses $m_1$ and $m_2$ respectively. As the external fermions are on
mass shell, we also have that $p_1^2 = m_1^2$, $p_2^2 = m_2^2$. As
such, the amplitude for the decay can can be written as $e
\epsilon_\mu^*(q) {\cal M}^\mu$, where $\epsilon_\mu^*(q)$ is the
polarization vector of the emitted photon. The most general ${\cal
M}^\mu$ for an on-shell photon\footnote{By an on-shell photon we mean
$\epsilon_\mu^*(q) q_\mu = 0$.} can be written as
\cite{Lavoura:2003xp}:  
\beq
{\cal M}^\mu =
i \bar{u}_2 \left[ \sigma^{\mu \nu} q_\nu \left(\sigma_L P_L + \sigma_R
P_R \right) \right] u_1 , 
\label{eq:13}
\eeq
where $P_L = (1 - \gamma_5)/2$, $P_R = (1 + \gamma_5)/2$ and
$\sigma_L$, $\sigma_R$ are the respective coefficients. From the
expression given in equation (\ref{eq:13}) we get the partial decay
width for \foftga as:  
\beq
\Gamma = \frac{\left(m_1^2 - m_2^2\right)^3}{16 \pi m_1^3}
\left(|\sigma_L|^2 + |\sigma_R|^2\right) . 
\label{eq:14}
\eeq

\par Assuming the fermions $f_1$ and $f_2$ interact with a neutral or
charged vector boson, $B_\alpha$, and with another fermion $F$, the
gauge interaction part of the Langrangian can be written as: 
\beq
{\cal L} = \sum_{i=1}^2 
\left[ B_\mu \bar{F} \gamma^\mu \left(L_i P_L + R_i P_R\right) f_i + 
B_\mu^* \bar{f}_i \gamma^\mu \left(L_i^* P_L + R^*_i P_R \right) F
\right] , 
\label{eq:15}
\eeq
where $L_i$ and $R_i$ are coefficients of the operators (model
dependent). In our case (LHT), from Table \ref{table:1} in Appendix
\ref{FeynRules}, we can see that we do not have any right-handed
currents contributing to the process. As such, $R_i = 0$. Equation
(\ref{eq:15}) then becomes, for our case: 
\beq
{\cal L} = \sum_{i=1}^2 
\left[ B_\mu \bar{F} \gamma^\mu L_i P_L f_i + 
B_\mu^* \bar{f}_i \gamma^\mu L^*_i P_L F
\right] . 
\label{eq:16}
\eeq
Note that we shall again assume that the Higgs exchange diagrams will
not contribute, as they are higher order in $(v/f)^2$.  

\begin{figure}[htb]
\bcen
\epsfig{file=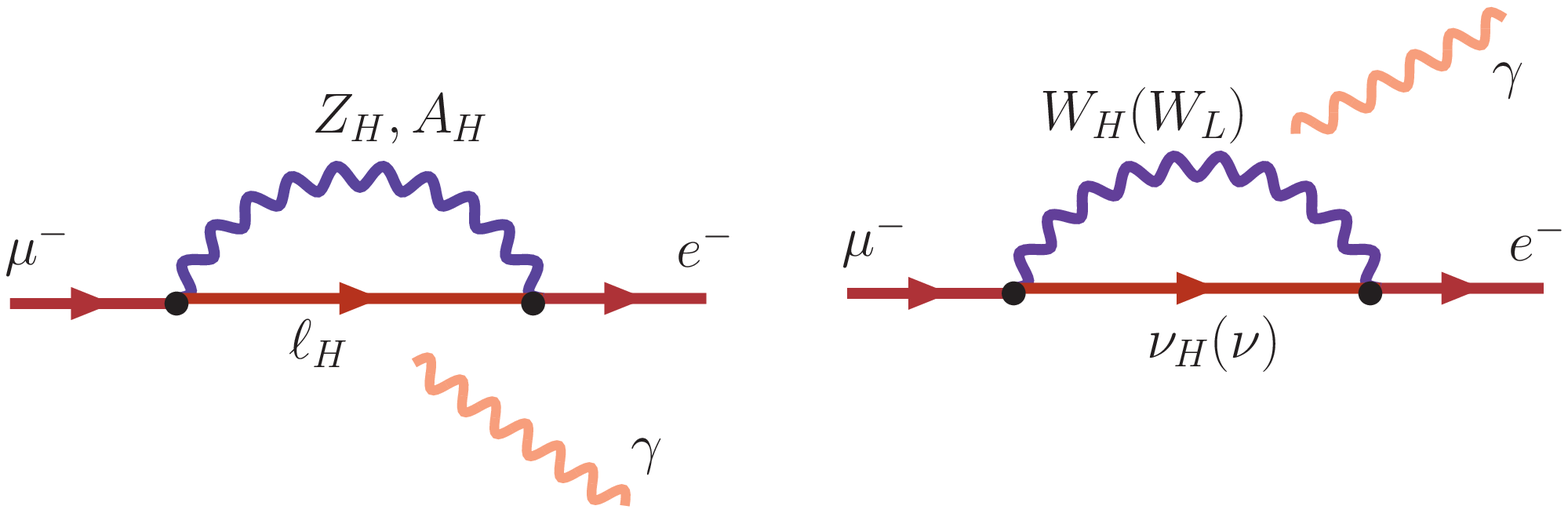,width=.8\textwidth}
\caption{The Feynman diagrams for \mutoeg in LHT}
\label{fig:mu2eg:1}
\ecen
\end{figure}

\noindent In the LH model we will also have contributions from loops
containing mirror fermions, $W_H, Z_H$ and $A_H$. Their contributions
to $\sigma_L$ and $\sigma_R$ (as defined in equation (\ref{eq:13}))
can be defined as: 
\beqa
\sigma_L = (\sigma_L)_{W_H} +  (\sigma_L)_{Z_H} + (\sigma_L)_{A_H} , 
\nonumber \\
\sigma_R = (\sigma_R)_{W_H} +  (\sigma_R)_{Z_H} + (\sigma_R)_{A_H} . 
\label{eq:17}
\eeqa
Where the expressions for the $\sigma$'s above are given in Appendix
\ref{LepFlavFunc}. However, as one final note, in our case (with
T-parity) equation (\ref{eq:15}) can be expressed as, using the
vertices given in Table \ref{table:1}:  
\beqa
{\cal L} &=& \frac{i g}{\sqrt{2}} \sum_{ij} \bar{\nu}_{H_i} \gamma_\mu
P_L (V_{H\ell})_{ij} \ell_j W_H^\mu 
+ i \left( - \frac{g}{\sqrt{2}} + \frac{g'}{10} x_H \frac{v^2}{f^2} \right) 
\sum_{ij} \bar{\ell}_{H_i} \gamma_\mu P_L (V_{H\ell})_{ij} \ell_j
Z_H^\mu                                \nonumber \\
&& + i \left( \frac{g'}{10} + \frac{g'}{10} x_H \frac{v^2}{f^2} \right) 
\sum_{ij} \bar{\ell}_{H_i} \gamma_\mu P_L (V_{H\ell})_{ij} \ell_j
A_H^\mu  + {\cal O} \left( \frac{v^4}{f^4} \right) .
\label{eq:18}
\eeqa

\par In the next section we shall study how the branching ratios for
the decays \mutoeg and \tautomug change for different values of the
parameter $f$, and for various mirror lepton masses. 


\section{Numerical Analysis \& Discussion } 

\par For our numerical results we shall use the standard
parameterisation the of $\vpmns$ matrix, which can be written as:  
\beqa
\vpmns &=& 
\left( \begin{array}{c c c}
       1 &  0         &  0        \\
       0 &    c_{atm}  &  s_{atm}   \\
       0 &  - s_{atm}  &  c_{atm}   
       \end{array}
\right)
\left( \begin{array}{c c c}
       c_{rct}                &  0  &  s_{rct} e^{-i \delta_r}   \\
       0                     &  1  &  0                      \\
       - s_{rct} e^{i \delta_r} &  0  &  c_{rct}   
       \end{array}
\right)
\left( \begin{array}{c c c}
       c_{sol}   &  s_{sol}    &  0        \\
       - s_{sol} &  c_{sol}    &  0        \\
       0        &  0         &  1
       \end{array}
\right)          \nonumber \\
&=& U(\theta_{atm}) U(\theta_{rct}) U(\theta_{sol}) , 
\nonumber 
\eeqa
where the $\theta_{atm}$, $\theta_{rct}$ and $\theta_{sol}$ are the
atmospheric, reactor and solar mixing angles. From neutrino
oscillation experiments we have $\triangle m_{12}^2 \sim 8 \times
10^{-5} {\rm eV^{-2}}$, $sin^2 2 \theta_{sol} \sim 0.31$, $|\triangle
m_{13}^2| \sim 2.6 \times 10^{-3} {\rm eV^2}$, $sin^2 2 \theta_{atm}
\sim 1.0$ and $sin\theta_{rct} \le 0.2$ (for our calculations we have
taken $\sin \theta_{rct} = 0.2$). From WMAP constraints we also have
that $\sum_{i=1,2,3} m_i < 2 eV$ (that is, the sum of the masses of
the three SM neutrino species). As such, the structure of the leptonic
sector mixing matrix, $\vpmns$, which is analogous to the quark sector
CKM matrix, can give rise to the lepton flavour violating processes
(such as \mutoeg, \tautomug, $\tau \to \mu (\pi,K)$, $\tau^- \to \mu^-
(e^-) \mu^+ \mu^-$ {\it etc.}) within the SM. Where, to reiterate,
these flavour violating processes, within the SM, will be dependent
upon the structure of the mixing matrix ($\vpmns$) and the neutrino
masses. Note that the smallness of neutrino mass, as indicated by WMAP
data, ensures the suppression of these processes to a level which
cannot be probed even in foreseeable future. Furthermore, the lepton
sector GIM mechanism suppresses the branching ratio of \mutoeg to a
value less than $10^{-40}$ within the SM. As such, we shall refer to
this situation as Minimal Flavour Violation (MFV).  

\par As discussed earlier, in LHT we can have a new mechanism for
lepton flavour violation, where, in the T-parity model the LFV can
come from the flavour mixing in the mirror fermion sector. The mixing
in that mirror fermion sector can, furthermore, give rise to a {\it
TeV scale GIM mechanism}. Note that this has been extensively
discussed in the case of hadronic decays \cite{Blanke:2006eb}. The
possible implications of this TeV scale GIM mechanism, in the case of
lepton sector, has been stressed in the T-parity model
\cite{Goyal:2006vq}. There is a possibility of large enhancement of
LFV decays in the T-parity model, despite the presence of a TeV scale
GIM mechanism in the lepton sector. As such, we shall quantify this by
calculating some definite values of the mixing matrix and other LH
parameters.  

\par The new mixing matrix which gives rise to flavour violation in
the lepton sector ($V_{H\ell}$), in general, has four parameters;
namely three angles and one phase. The presence of this mixing matrix
arises from the possibility of a departure from MFV, which was present
within the SM. We therefore parameterise this mixing matrix with three
mixing angles ($\theta_{12}, \theta_{23}, \theta_{13}$) and a phase
($\delta$) as:  
\beqa
V_{H\ell} &=& 
\left( \begin{array}{c c c}
       1 &  0         &  0        \\
       0 &    c_{23}  &  s_{23}   \\
       0 &  - s_{23}  &  c_{23}   
       \end{array}
\right)
\left( \begin{array}{c c c}
       c_{13}                &  0  &  s_{13} e^{-i \delta}   \\
       0                     &  1  &  0                      \\
       - s_{13} e^{i \delta} &  0  &  c_{13}   
       \end{array}
\right)
\left( \begin{array}{c c c}
       c_{12}   &  s_{12}    &  0        \\
       - s_{12} &  c_{12}    &  0        \\
       0        &  0         &  1
       \end{array}
\right)          \nonumber \\
&=& U(\theta_{23}) U(\theta_{13}) U(\theta_{12}) . 
\nonumber 
\eeqa
\noindent At this point we would like to stress that for the departure
from MFV, within the SM, the following conditions need to be
satisfied: 
\begin{enumerate}
\item[(1)] The matrix $V_{H\ell}$ should be different from the Identity matrix.
\item[(2)] And that the three generations of mirror fermions should
not be degenerate in mass.  
\end{enumerate}
\noindent From the above structural considerations of the mixing
matrices, where as a benchmark we will show the results of LFV
processes for the following cases:  
\begin{enumerate}
\item[] {\bf Case A}: Where we assume that $V_{H\ell}$ is related to
$\vpmns$, such that no additional new parameters, except those of the
masses of mirror fermions, will be needed. 
\item[] {\bf Case B}: Where we assume the hierarchy of the mixing
angles to be: 
\beq
s_{12} \ll s_{13} \ll s_{23} . 
\eeq
\item[] {\bf Case C}: And finally, where we assume that the hierarchy
of the mixing angles is: 
\beq
s_{12} \ll s_{23} \ll s_{13} . 
\eeq
\end{enumerate}
\noindent We shall now analyze the effects of the above cases on our
observables for \gmtwo, \mutoeg and \tautomug, where the present
experimental bounds for the observables are\cite{Yao:2006px}:  
\beqa
\delta a_\mu &=& 22(10) \times 10^{-10} ,  \nonumber \\
Br(\mu \to e \gamma)  &\le& 1.2 \times 10^{-11} \quad [90 \% \ C.L.],
\nonumber \\ 
Br(\tau \to \mu \gamma) &\le& 6.8 \times 10^{-8} \quad [90 \% \ C.L.] . 
\nonumber 
\eeqa

\par
\begin{figure}[htb]
\bcen
\epsfig{file=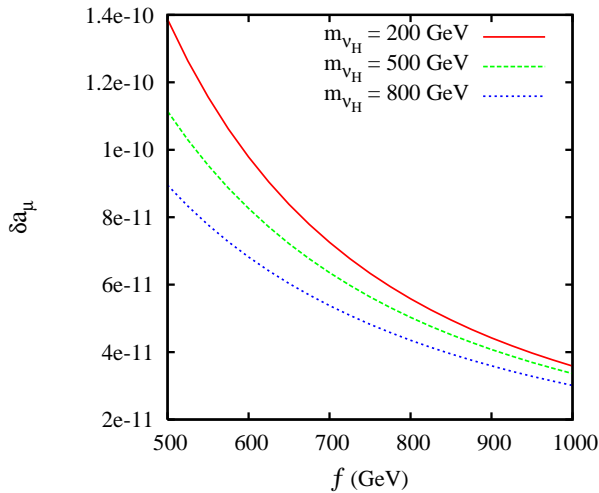,width=.6\textwidth}
\caption{\it $\delta a_{\mu}$, where we have assumed $m_{\ell_H} =
m_{\nu_H}$, i.e. the same mass for all the mirror fermion doublets} 
\label{fig:2}
\ecen
\end{figure}

 Firstly, we shall present the results for $\delta a_\mu$ in the
case where the mirror leptons have degenerate mass. In this case we
only have two parameters in the model, namely; the LH scale ($f$) and
mass of the mirror leptons. The results are plotted in Fig
(\ref{fig:2}). As mentioned above, this is the MFV case of the SM, and
hence \mutoeg and \tautomug will stay at their SM levels. Note that,
as we can be observed from the graph, although $\delta a_\mu$ shows
substantial variation as a function of the LH scale, its value is
still much lower than the present experimental bounds.  

\subsection{Case A}

\par In this case we shall consider $V_{H\ell}$ as being related to
$\vpmns$. In this case we do not introduce any additional parameters,
except the masses of the mirror fermions. For $\vpmns$ we take the
standard parameterisation with parameters given by the neutrino
experiments. Furthermore, we shall discuss the four cases, in analogy
to the discussion given in reference \cite{Akeroyd:2006bb}, namely: 
\begin{enumerate}
\item[{\bf I}]:  $V_{H\ell} = V_{PMNS}$ . 
\item[{\bf II}]:  $V_{H\ell} = U(\theta_{atm}) U(\theta_{rct})$ . 
\item[{\bf III}]:  $V_{H\ell} = U(\theta_{atm})$ . 
\item[{\bf IV}]:  $V_{H\ell} = I$ . 
\end{enumerate}

\begin{figure}[htb]
\bcen
\epsfig{file=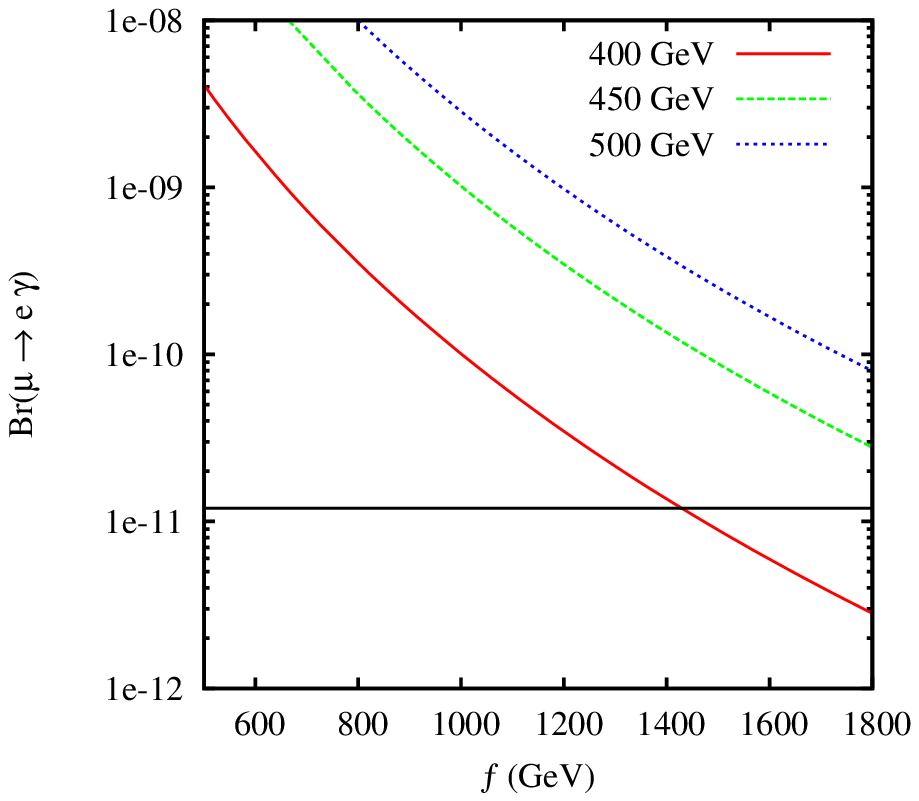,width=.5\textwidth} \hskip -1cm
\epsfig{file=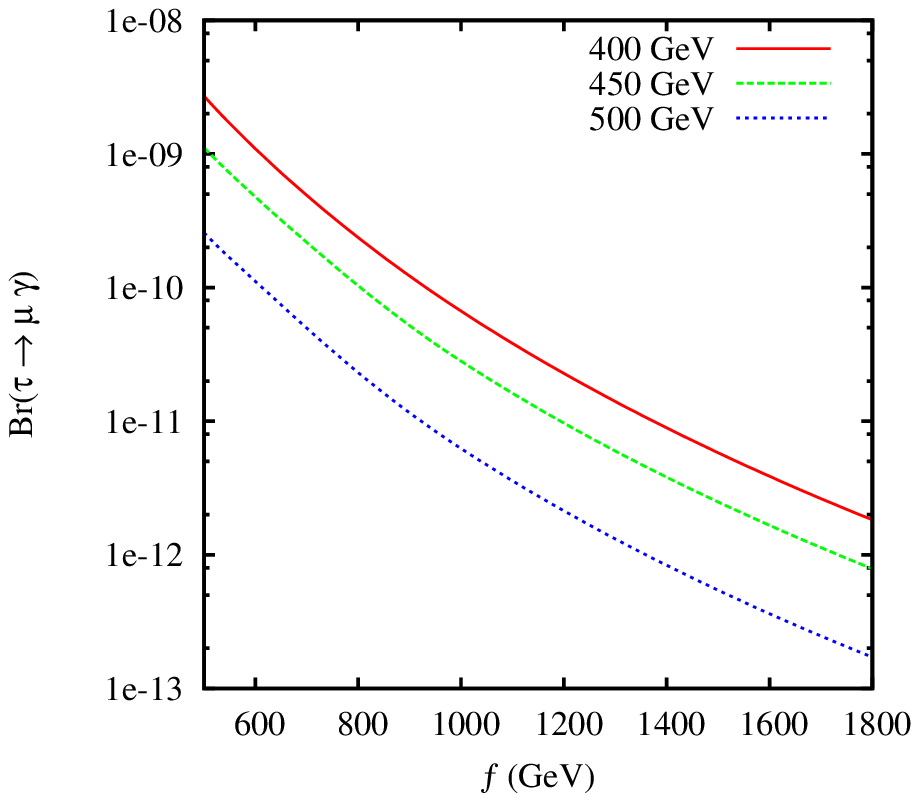,width=.5\textwidth}
\caption{\sl {\bf Case A - I} ($V_{H\ell} = V_{PMNS}$): \mutoeg
(\tautomug) as a function of $f$ in the left (right) panel for
different values of the second generation mirror lepton mass. The
masses of first and third generation mirror leptons are 400 GeV and
500 GeV respectively.}  
\label{fig:mu2eg:2}
\ecen
\end{figure}

\begin{figure}[htb]
\bcen
\epsfig{file=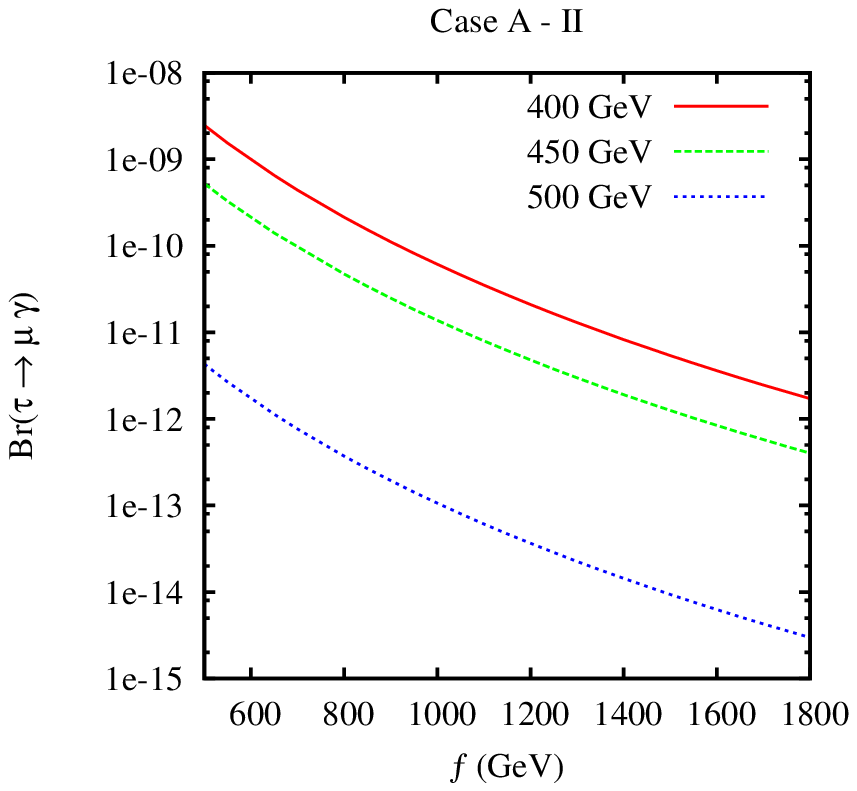,width=.57\textwidth} \hskip -3cm
\epsfig{file=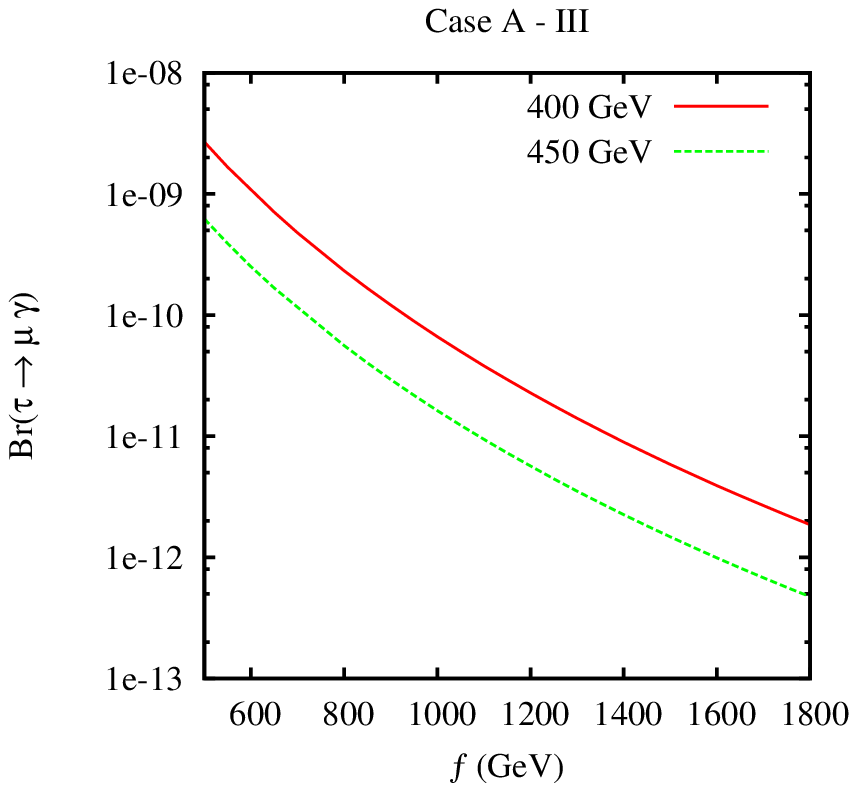,width=.57\textwidth}
\caption{\sl {\bf Case A - II (left panel), III (right panel)}: The
\tautomug decay as a function of $f$ for different values of the
second generation mirror lepton mass. The masses of the first and
third generation of the mirror leptons are 400 GeV and 500 GeV
respectively.}   
\label{fig:mu2eg:23}
\ecen
\end{figure}

\noindent{\bf Case A - I :} We have presented our results of case {\bf
I} in Fig (\ref{fig:mu2eg:2}). As can be seen from the figure, in this
case the branching ratios are very sensitive to the mass splitting of
the mirror leptons. Furthermore, in this case the experimental
measurement of \mutoeg practically rules out any substantial mass
splitting between all three generations of the mirror leptons. For
this case we have also shown a scatter plot of the correlation between
\mutoeg and \tautomug in Fig (\ref{fig:mu2eg:3}). In this plot we have
varied the masses of the mirror leptons in the range 500-600 GeV. As
can be seen from this figure, the \mutoeg decay practically rules out
most of the region where there is splitting between the mirror lepton
masses. However, there are still some regions where we can have a
fairly high (although still within the experimental limits) rate for
the \tautomug decay. Note that any improvement in the \tautomug decay
rate in the future would further constrain these parameters. 
\begin{figure}[htb]
\bcen
\epsfig{file=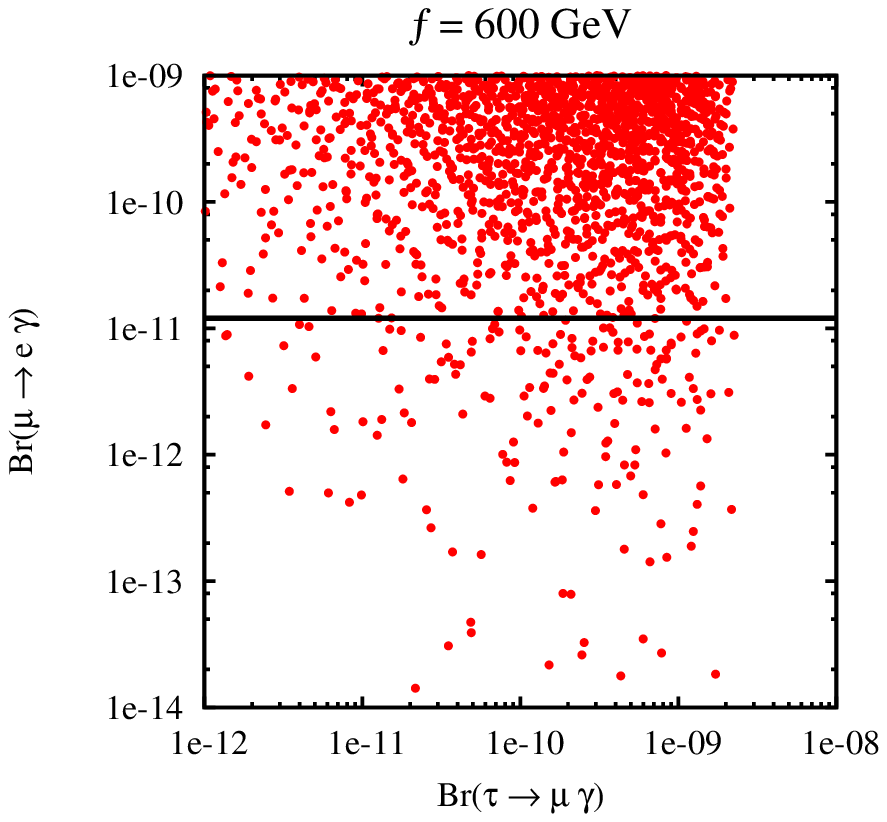,width=.55\textwidth} \hskip -2cm
\epsfig{file=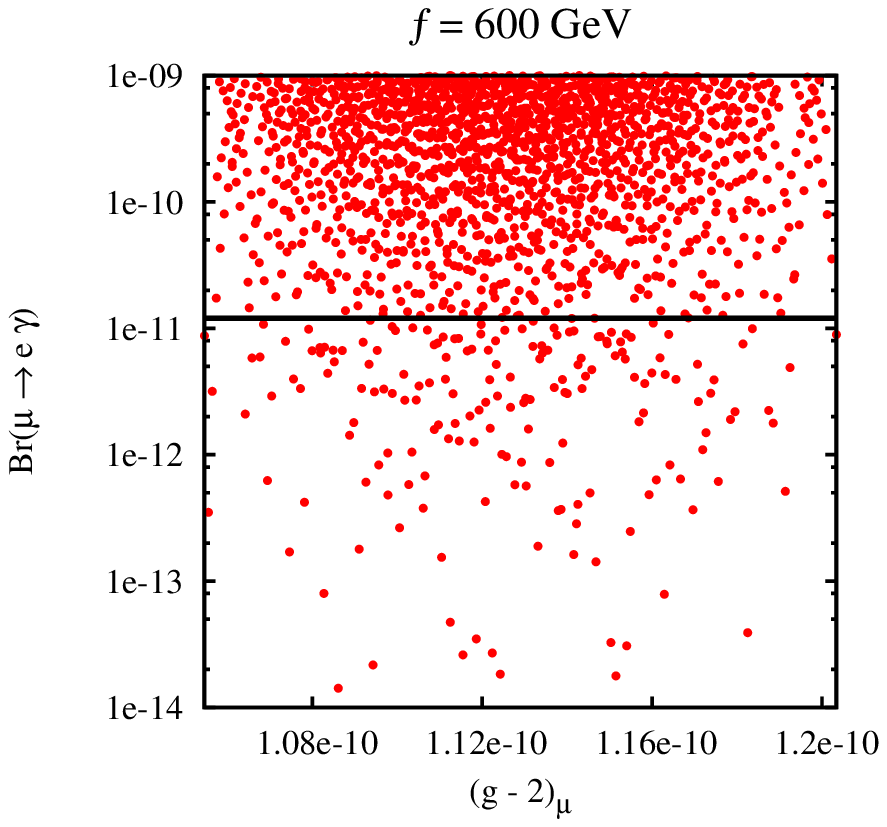,width=.55\textwidth}
\caption{\sl {\bf Case A} Co-relation between the $\tau \to \mu
\gamma$ and $\mu \to e \gamma$ decays (left panel) and the $\mu \to e
\gamma$  and $(g-2)_\mu$ (right panel). Masses of the heavy (mirror)
leptons are varied randomly in the range $500 < m_{\ell_H} < 700$ GeV
and $f =  600$ GeV.} 
\label{fig:mu2eg:22}
\ecen
\end{figure}
\vskip .4cm
\noindent{\bf Case A - II \& III :} In these cases there is no
significant change in the rate of the \mutoeg decay for the range of
mirror lepton masses considered here. The reason for this is due to
these cases corresponding to the the situation where $V_{H\ell} =
V_{PMNS}$, and $s_{12} = 0$ for case II, and $s_{12} = s_{13} = 0$ for
case III. In this case we do not have any appreciable mixing in the
first two generations of the mirror leptons, and hence no great change
in the predictions of the \mutoeg decay. However, the rate of the
\tautomug decay can be substantially changed. We have plotted the rate
of the \tautomug decay as a function of LH scale for case II and III
in Fig (\ref{fig:mu2eg:23}). As can be seen from these plots in Fig
(\ref{fig:mu2eg:23}), the predictions are still below the present
experimental bounds. However, if data for the \tautomug decay were
improved in future, from high luminosity SuperB factories, this would
help to constrain the possibilities greatly.  

\vskip .4cm
\noindent{\bf Case A - IV :} This is the MFV limit of
the LHT model. In this case as there is no mixing in mirror lepton
sector hence there will not be any contribution to LFV processes. 

\subsection{Case B}

\par
\begin{figure}[htb]
\bcen
\epsfig{file=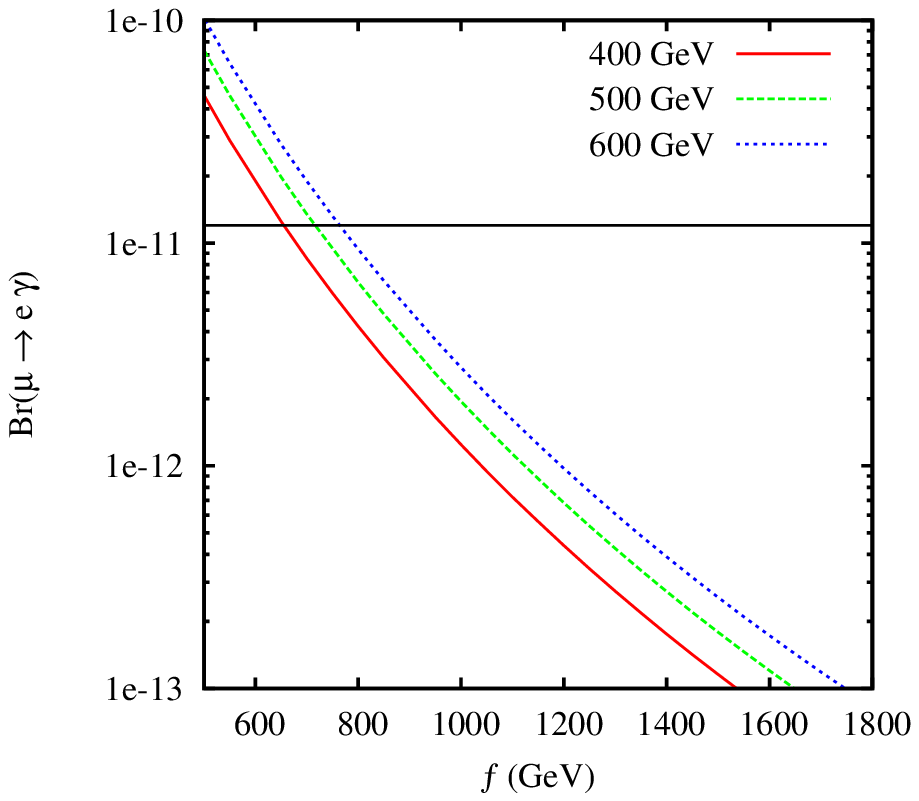,width=.55\textwidth} \hskip -2cm
\epsfig{file=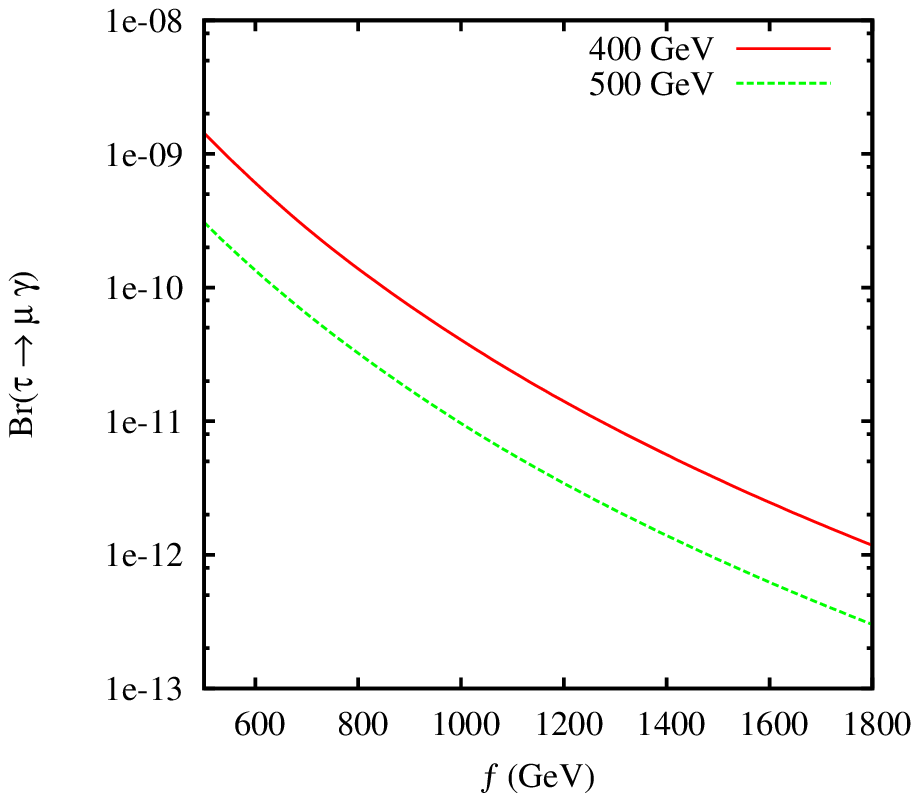,width=.55\textwidth}
\caption{\sl {\bf Case B}: The \mutoeg (\tautomug) decay as a function
of $f$, in the left (right) panel, for different values of the second
generation mirror lepton mass. The masses of the first and third
generation of mirror leptons are 400 GeV and 600 GeV respectively.}  
\label{fig:mu2eg:3}
\ecen
\end{figure}

 In this case we have assumed the pattern $s_{12} \ll s_{13} \ll
s_{23}$. In Fig (\ref{fig:mu2eg:3}) we have shown the results for the
fixed values of the angles, as given by: 
$$s_{23} \sim 0.2 , ~~ s_{13} \sim 0.02, ~~ s_{12} \sim 0.002 . $$
This pattern assures a very small mixing in the first two generations
of the mirror leptons, which ensures we keep the \mutoeg decay rate
rather low. However, in this case we can still have sufficient mixing
in the second and third generations to have a higher rate (although
still within the present experimental bounds) for the \tautomug decay.

\subsection{Case C}

\par
\begin{figure}[htb]
\bcen
\epsfig{file=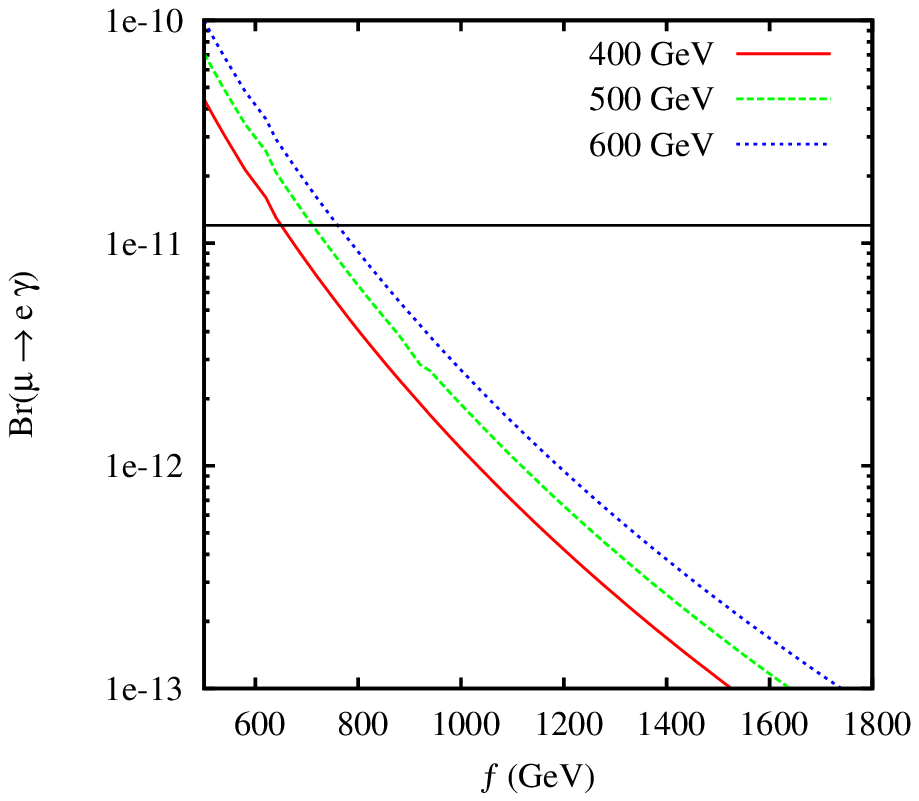,width=.55\textwidth} \hskip -2cm
\epsfig{file=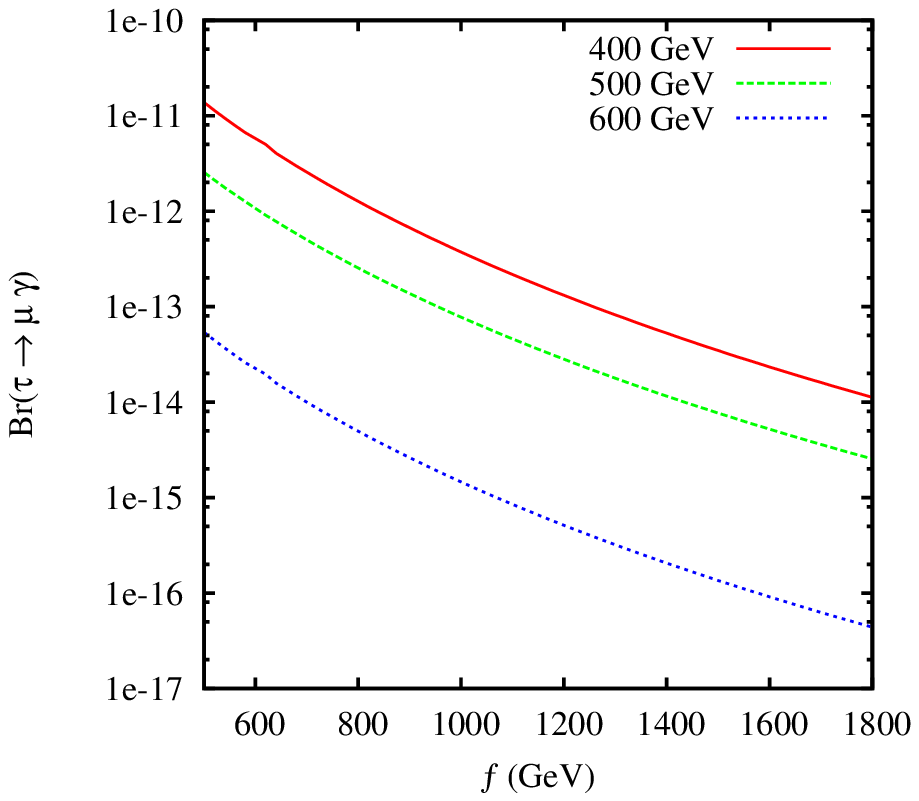,width=.55\textwidth}
\caption{\sl {\bf Case C}: The \mutoeg (\tautomug) decay as a function
of $f$, in the left (right) panel, for different values of the second
generation mirror lepton mass. The masses of the first and third
generation of mirror leptons are 400 GeV and 600 GeV respectively.}  
\label{fig:mu2eg:4}
\ecen
\end{figure}

 In this case we are assuming a hierarchy $s_{12} \ll s_{23} \ll
s_{13}$, where in Fig (\ref{fig:mu2eg:4}) we have plotted for specific
values of the angles, given by:  
$$s_{23} \sim 0.02, \quad
s_{13} \sim 0.2, \quad
s_{12} \sim 0.002 . 
$$
In this case the mixing matrix $V_{H\ell}$ is essentially diagonal. As
such there is very little mixing between the mirror leptons, which
results in a lower value for the decay rates of \mutoeg and
\tautomug. 

\par To summarize, the precision constraints on the anomalous magnetic
moment of muon ($\delta a_\mu$) does not constrain the LHT. However,
as T-parity models have new source of lepton flavour violation one can
extract useful constraints on model parameters from various lepton
flavour violating processes. In this paper we have analyzed the
effects of these new flavour violations in T-parity models on the
decays \mutoeg and \tautomug. From the experimental results of the
\mutoeg decay, rather strong constraints on the texture of the new
flavour violating mixing matrix, and the mass splitting of mirror
leptons is given. Note that there are practically no constraints from
the \tautomug decay on the model parameters from present experimental
results, however, the future SuperB factories, which may observe the
\tautomug and $\tau \to e \gamma$ decays at rates of $10^{-10}$, might
provide us very useful constraints on the model parameters. 


\section*{Acknowledgments}
NG would like to thank Yasuhiro Okada and Mayumi Aoki for
discussions. We would also like to thank J. Hubisz for his useful 
clarifications on the T-parity model. The work of NG was supported by
the JSPS, under fellowship no P06043. The work of SRC is supported by
Ramanna fellowship of DST, India.  


\appendix

\section{Feynman Rules}\label{FeynRules}

\par In this appendix we list all the relevant Feynman rules for our
analysis, which have been summarized in Table \ref{table:1}
\cite{Blanke:2006eb}.  

\begin{table}[htb]
\bcen
\begin{tabular}{| c | c || c | c |}  \hline
Particles &  Vertices  & Particles & Vertices \\ \hline
$\bar{\ell}_i W_L^{-\mu} \nu_j $ & 
    $i \frac{g}{\sqrt{2}} \left(V_{PMNS}\right)_{ij} P_L $ & 
$\bar{\nu}_{H_i} W_H^{+\mu} \ell_j$ & 
    $i \frac{g}{\sqrt{2}} \left(V_{H\ell}\right)_{ij} P_L $ \\ \hline 
$\bar{\ell}_i Z_L^\mu \ell_j$ &  
  $\frac{i g}{cos\theta_w} \gamma^\mu \left[ \left(- {1 \over 2} +
      sin^2\theta_w\right) P_L + sin^2\theta_w P_R \right] \delta_{ij}$ &
$\bar{\ell}_{H_i} Z_H^\mu \ell_j$ &  
    $i \left(- \frac{g}{2} + \frac{ g'}{10} x_H \frac{v^2}{f^2} \right)
    \left(V_{H_\ell}\right)_{ij} \gamma^\mu P_L$  \\ \hline
$\bar{\ell}_i A_L^\mu \ell_j$ &  
  $- i g' \gamma^\mu \delta_{ij}$ &
$\bar{\ell}_{H_i} A_H^\mu \ell_j$ &  
   $ i \left(\frac{g'}{10} + \frac{g'}{10} x_H \frac{v^2}{f^2} \right) 
   \left(V_{H_\ell}\right)_{ij} \gamma^\mu P_L$
  \\ \hline
\end{tabular}
\caption{where $x_H = \frac{5 g g'}{4 (5 g^2 -
g'^2)}$. Vertices taken from Buras {\it et al.}\cite{Blanke:2006eb} }
\label{table:1}
\ecen
\end{table}

\section{Functions for \gmtwo}\label{GminusTwo}

\par The functions used in the determination of the LH contribution to
\gmtwo are \cite{Leveille:1977rc}: 
\beqa
F_{W_H}(x_i) &=& \int^1_0 dy
\frac{- 2 y^2 (1 + y) - x_i (2 y - 3 y^2 + y^3) - x_\mu y^2 (y - 1)}{y
  + x_\mu (y^2 - y) + x_i (1 - y)},   
\label{eq:8} \\
F_{Z_H}(y_i) &=& \int^1_0 dx 
\frac{(x - x^2)(x-2) - {1 \over 2} \left( y_i (x^3 + x^2) + x_\mu (x^3 - x^2)
  \right)}{(1 - x) + x_\mu (x^2 - x) + y_i x} , 
\label{eq:9}
\eeqa
with $x_\mu = \left(\frac{m_\mu}{M_{W_H}}\right)^2$.

\par In the limit $x_\mu \to 0$, that is, where we neglect the $\mu$
mass when compared to $M_{W_L}$, the above integrations can be
analytically expressed as: 
\beqa
F_{W_H}(x_i) &=& 
- \frac{10 - 43 x_i + 78 x_i^2 - 49 x_i^3 + 4x_i^4 + 18x_i^3Log(x_i) }
{6\left( x_i - 1 \right)^4} , \label{eq:10} \\
F_{Z_H}(y_i) &=& 
\frac{- 8 + 38 y_i - 39 y_i^2 + 14 y_i^3 - 5 y_i^4 + 18 y_i^2 Log(y_i)
}{12(y_i - 1)^4} . 
\label{eq:11}
\eeqa

\section{Functions for \mutoeg and \tautomug}\label{LepFlavFunc}

\par In determining the branching ratio for the decays \mutoeg and
\tautomug, we have made use of the following expressions
\cite{Lavoura:2003xp}. Firstly note that apon comparing equation
(\ref{eq:16}) with equation (\ref{eq:17}) for the \mutoeg decay the
coefficients $L_i$ will have values: 
\beq
\begin{array}{l  l  l }
W_H \ contribution :  &
  (L_\mu^{W_H})^*_i = 
       \frac{i g}{\sqrt{2}} \left(V_{H\ell}\right)_{2i}, &
  (L_e^{W_H})^*_i$ =  
    - $\frac{i g}{\sqrt{2}} \left(V_{H\ell}\right)^*_{i1}       \\
Z_H \ contribution : &
  (L_\mu^{Z_H})_i = i \left( - \frac{g}{\sqrt{2}} + \frac{g'}{10} x_H
          \frac{v^2}{f^2} \right) \left(V_{H\ell}\right)_{2i}, &
  (L_e^{Z_H })^*_i = - i \left( - \frac{g}{\sqrt{2}} + \frac{g'}{10} x_H
          \frac{v^2}{f^2} \right) \left(V_{H\ell}\right)^*_{i1}  \\
A_H \ contribution : &  
  (L_\mu^{A_H})_i = i \left( \frac{g'}{10} + \frac{g'}{10} x_H
          \frac{v^2}{f^2} \right) \left(V_{H\ell}\right)_{2i}, &
  (L_e^{A_H})^*_i = - i \left( \frac{g'}{10} + \frac{g'}{10} x_H
          \frac{v^2}{f^2} \right) \left(V_{H\ell}\right)^*_{i1} , \\ 
\end{array}
\label{eq:19}
\eeq
where the index $i$ sums over the three generation of mirror fermions,
where analogous expressions for the decay \tautomug are given by
selecting the appropriate elements in $V_{H\ell}$. Defining now the
product: 
\beq
\lambda^X_i = (L_\mu^X)_i  (L_e^X)^*_i , \quad X = W_H, Z_H, A_H,
\label{eq:20}
\eeq
the LH contributions to $\sigma$, as defined in equation (\ref{eq:17}), can be written as:
\beq
\begin{array}{l l l}
W_H \ loop :~~ & (\sigma_L)_{W_H} = Q_{W_H} \ \lambda^{W_H}_i \ \bar{y}_2(m_{\nu_H}^i,m_{W_H}), ~~~  & 
(\sigma_R)_{W_H} = Q_{W_H} \ \lambda^{W_H}_i \ \bar{y}_1(m_{\nu_H}^i,m_{W_H}) ,  \\
Z_H \ loop :~~ & (\sigma_L)_{Z_H} = Q_{\ell_H} \ \lambda^{Z_H}_i \ y_2(m_{\ell_H}^i,m_{Z_H}), ~~~  & 
(\sigma_R)_{Z_H} = Q_{\ell_H} ~ \lambda^{Z_H}_i \ y_1(m_{\ell_H}^i,m_{Z_H}) ,  \\
A_H \ loop :~~ & (\sigma_L)_{A_H} = Q_{\ell_H} \ \lambda^{A_H}_i \ y_2(m_{\ell_H}^i,m_{A_H}), ~~~  & 
(\sigma_R)_{A_H} = Q_{\ell_H} ~ \lambda^{A_H}_i \ y_1(m_{\ell_H}^i,m_{A_H}) , \\
\end{array}
\label{eq:21}
\eeq
where $Q_{W_H}$ is the charge of $W_H$ and $Q_{\ell_H}$ is the charge
of heavy mirror lepton. The loops and y functions are given as: 
\beqa
y_1(m_F,m_B) &=& m_1 \left[ 2 a + 4 c_1 + 2 c_2 + 2 d_1 + 2 f +
\frac{m_F^2}{m_B^2} \left(- c_2 + d_1 + f\right) + \frac{m_2^2}{m_B^2}
\left(c_2 + d_2 + f\right)\right] ,        \nonumber \\
y_2(m_F,m_B) &=& m_2 \left[ 2 a + 2 c_1 + 4 c_2 + 2 d_1 + 2 f +
\frac{m_F^2}{m_B^2} \left(- c_1 + d_2 + f\right) + \frac{m_1^2}{m_B^2}
\left(c_1 + d_1 + f\right)\right] ,        \nonumber \\
\bar{y}_1(m_F,m_B) &=& m_1 \left[ 2 \bar{c}_2 + 2 \bar{d}_1 + 2 \bar{f} +
\frac{m_F^2}{m_B^2} \left( \bar{a} - 2 \bar{c}_1 - \bar{c}_2 +
\bar{d}_1 + \bar{f}\right) + \frac{m_2^2}{m_B^2} \left(- \bar{c}_2 +
\bar{d}_2 + \bar{f}\right)\right] ,      \nonumber \\
\bar{y}_2(m_F,m_B) &=& m_2 \left[ 2 \bar{c}_1 + 2 \bar{d}_2 + 2 \bar{f} +
\frac{m_F^2}{m_B^2} \left( \bar{a} - \bar{c}_1 - 2 \bar{c}_2 +
\bar{d}_2 + \bar{f}\right) + \frac{m_1^2}{m_B^2} \left(- \bar{c}_1 +
\bar{d}_1 + \bar{f}\right)\right] . 
\label{eq:22}
\eeqa
Where in the above equations we have used:
\beqa
a &=& \frac{i}{16 \pi^2} C_0(m_1^2,q^2,m_2^2,m_B^2,m_F^2,m_F^2) ,
     \nonumber \\
c_1 &=& \frac{i}{16 \pi^2} C_1(m_1^2,q^2,m_2^2,m_B^2,m_F^2,m_F^2) ,
     \nonumber \\
c_2 &=& \frac{i}{16 \pi^2} C_2(m_1^2,q^2,m_2^2,m_B^2,m_F^2,m_F^2) ,
     \nonumber \\
d_1 &=& \frac{i}{16 \pi^2} C_{11}(m_1^2,q^2,m_2^2,m_B^2,m_F^2,m_F^2) ,
     \nonumber \\
d_2 &=& \frac{i}{16 \pi^2} C_{22}(m_1^2,q^2,m_2^2,m_B^2,m_F^2,m_F^2) ,
     \nonumber \\
f &=& \frac{i}{16 \pi^2} C_{12}(m_1^2,q^2,m_2^2,m_B^2,m_F^2,m_F^2) ,
\label{eq:23}
\eeqa
and where $C_0$, $C_1$, $C_2$, $C_{11}$, $C_{22}$ and $C_{12}$ are the
PV functions. If we now use the approximation that $m_1^2 = m_2^2 = 0$
and $q^2 = 0$ the above PV functions can be written in terms of $t =
m_F^2/m_B^2$ as \cite{Lavoura:2003xp}:  
\beqa
a &=& \frac{i}{16 \pi^2 m_B^2} 
    \left[ \frac{-1}{t - 1} + \frac{ln t}{(t - 1)^2}\right],
         \nonumber \\
c_1 = c_2 = c &=& \frac{i}{16 \pi^2 m_B^2}
    \left[ \frac{t - 3}{4 (t - 1)^2} + \frac{ln t}{2 (t - 1)^3}
          \right] ,    \nonumber \\
d_1 = d_2 = 2 f = d &=& \frac{i}{16 \pi^2 m_B^2}
    \left[ \frac{- 2 t^2 + 7 t - 11}{ 18 (t - 1)^3} + 
       \frac{ln t}{3 (t - 1)^4} \right],   \nonumber \\
\bar{a} &=& \frac{i}{16 \pi^2 m_B^2} 
    \left[ \frac{1}{t - 1} - \frac{t ln t}{(t - 1)^2} 
        \right],        \nonumber \\
\bar{c}_1 = \bar{c}_2 = \bar{c} &=& \frac{i}{16 \pi^2 m_B^2}
    \left[ \frac{3 t - 1}{4 (t - 1)^2} - \frac{t^2 ln t}{2 (t - 1)^3}
        \right] ,       \nonumber  \\
\bar{d}_1 = \bar{d}_2 = 2 \bar{f} = \bar{d} &=& \frac{i}{16 \pi^2 m_B^2}
    \left[ \frac{11 t^2 - 7 t + 2}{18 (t - 1)^3} - 
        \frac{t^3 ln t}{3 (t - 1)^4} \right] .      
\label{eq:24}
\eeqa

Using equation (\ref{eq:21}), and equation (\ref{eq:17}) in equation
(\ref{eq:14}), we can obtain the branching ratio.  


\end{document}